\documentclass[floats,aps,twocolumn,showpacs,amsmath,amssymb,prb]{revtex4}
\usepackage{graphicx}
\usepackage{color}
\usepackage{amsmath}
\usepackage{dcolumn}
\usepackage{bm}

\input epsf
\def\k{{\bf k}}

\def\q{{\bf q}}
\begin{document}
\newcommand{\ltwid}{\mathrel{\raise.3ex\hbox{$<$\kern-.75em\lower1ex\hbox{$\sim$}}}}
\newcommand{\gtwid}{\mathrel{\raise.3ex\hbox{$>$\kern-.75em\lower1ex\hbox{$\sim$}}}}
\newcommand{\BSCCO}{{Bi$_2$Sr$_2$CaCu$_2$O$_8$ }}

\title{Elastic forward scattering in the cuprate superconducting state }

\author{L. Zhu$^{1}$\email{zhu@phys.ufl.edu}, P.J. Hirschfeld$^1$, and D.J.~Scalapino$^{2}$\email{djs@vulcan2.physics.ucsb.edu}}

\affiliation{${^1}$Department of Physics, University of Florida,
Gainesville, FL 32611\\$^2$Department of Physics, University of
California, Santa Barbara, CA 93106}

\date{\today}

\begin{abstract}

We investigate the effect of elastic forward  scattering on the
ARPES spectrum of the cuprate superconductors. In the normal
state,  small angle scattering from out-of-plane impurities is
thought to broaden the ARPES spectral response with minimal effect
on the resistivity or the superconducting transition temperature
$T_c$. Here we explore how such forward scattering affects the
ARPES spectrum in the $d$-wave superconducting  state.  Away from
the nodal direction, the one-electron impurity scattering rate is
found to be suppressed as $\omega$ approaches the gap edge by a
cancellation between normal and anomalous scattering processes,
leading to a square-root-like feature in the spectral weight as
$\omega$ approaches $-\Delta_\k$ from below. For momenta away from
the Fermi surface, our analysis suggests that a dirty  optimally
or overdoped system will still display a sharp but nondispersive
peak which could be confused with a quasiparticle spectral
feature. Only in cleaner samples should the true dispersing
quasiparticle peak become visible. At the nodal point on the Fermi
surface, the contribution of the anomalous scattering vanishes and
the spectral weight exhibits a Lorentzian quasiparticle peak in
both energy and momentum.
 Our analysis, including a treatment of unitary scatterers and inelastic
 spin fluctuation scattering, suggests explanations for the
sometimes mysterious lineshapes and temperature dependences of the
 peak structures observed in the \BSCCO system.


\end{abstract}

 \maketitle



\section{Introduction}


Since the earliest ARPES studies of optimally doped \BSCCO
(BSCCO), spectral features near the Fermi level have been reported
whose width suggested the existence of a significant elastic
scattering contribution which varied along the Fermi surface,
taking its smallest value along the $(\pi, \pi)$ diagonal and
largest near the $(\pi, 0)$
region\cite{Wel95,Ron98,Val00,Kam00,DHS03}. The shapes of the
measured energy dispersion curves (EDCs) are quite unusual,
suggesting that the 1-electron self energy is also strongly energy
dependent. Abrahams and Varma\cite{AV00} attempted to account for
both features by assuming that the self energy was a sum of an
energy-independent part arising from small-angle (forward)
scattering, and a second, momentum independent term modelled with
marginal Fermi liquid theory.  They noted that the elastic forward
scattering was most probably associated with impurities which were
located away from the CuO$_2$ planes and reflected the $v_F({\bf
k})^{-1}$ variation of the momentum-resolved density of states as
$\k$ moved around the Fermi surface. They further remarked that,
despite the large scattering rates of order 100meV deduced from
fits to ARPES spectra, the forward scattering nature of the
disorder  would be consistent with such scattering having a
negligible effect on the resistivity of the optimally doped
cuprates\cite{Gia94}.  In addition, it was shown that this type of
disorder would have a small effect on $T_c$\cite{Kee01}.

There are two obvious difficulties with this scenario.  The first
is that the spectral peak measured by ARPES near the $(\pi,0)$
point is known to sharpen dramatically when one goes below $T_c$,
a phenomenon interpreted as the formation of a coherent
quasiparticle in the superconducting state.  This sharpening has
normally been attributed to the well-known collapse of the
inelastic scattering rate below $T_c$ due to the opening of the
superconducting gap, but it is hard {\it a priori} to guess why
something similar should happen in the presence of an elastic
scattering rate of order 100meV. The second problem is that
recently increased momentum resolution\cite{DHS03} and the use of
different photon energies\cite{Fen01,Chu01,Bog01,Kor02} has
resolved a bilayer splitting which has its maximum effect near the
$(\pi,0)$ point. Some of the previously observed ``elastic
broadening" is therefore certainly due to this as well as to
pseudogap effects\cite{Kam04}, but exactly how much is not clear.

On the other hand, one cannot ignore the out-of-plane disorder.
The BSCCO material is thought to be doped by excess oxygen in the
SrO and BiO planes, and even the best single crystals are believed
to contain significant amounts of cation switching and other
out-of-plane defects\cite{Eisaki}. It is therefore reasonable to
assume that quasiparticles moving in the CuO$_2$ planes of this
material must experience a smooth potential landscape due to these
defects,  and useful to pursue the question of the effect of this
type of scattering in the superconducting state.  In fact,
 fits\cite{ZAH} to Fourier transformed-scanning tunnelling spectroscopy
measurements\cite{McE03} on similar samples to those used in the
ARPES studies have recently been shown to require both a strong
(near-unitary limit) scattering component, attributed to native
defects in the Cu0$_2$ planes, as well as a weaker, smooth
scattering potential component attributed to defects away from the
plane. Recently, Markiewicz has also attempted to relate STM and
ARPES data in the superconducting state assuming a smooth
potential.\cite{Mar03}

In this work we model the complex collection of out-of-plane
defects with a simple set of impurity potentials with finite
range, and find a number of surprising results.  The first is that
the elastic scattering rate indeed ``collapses" in the
superconducting state in the strong forward scattering limit,
leading to a sharp spectral feature everywhere on the Fermi
surface except at the nodal point itself.  The second is that as
one goes away from the Fermi surface, this feature disperses away
as expected for a quasiparticle peak {\it only} if the system is
sufficiently clean; otherwise it remains pinned to the gap edge.
This would appear to explain the apparent lack of quasiparticle
dispersion in older samples.  In some situations both a
nondispersive gap edge peak and a dispersing quasiparticle feature
can be simultaneously observed. While we do not attempt a direct
fit to experiment, we show that assuming a rather simple and
physically motivated model for the one electron self-energy, which
combines the above description of forward elastic scattering with
strong pointlike elastic scattering  and spin fluctuation
inelastic scattering, allows us to calculate a spectral function
which appears to reproduce many of the qualitative features of
current ARPES data on optimally to overdoped cuprates.
 It is our
hope that the ideas presented here can help unravel some of the
mysteries surrounding the behavior of what is generally called the
superconducting ``quasiparticle peak", and allow a more accurate
description of the actual propagating excitation.  They will also
have important immediate implications for other bulk properties of
the superconducting state.

The paper is organized as follows.
 Section II  discusses general effects of extended impurities
scattering on the ARPES spectrum in both the normal and
superconducting states.  In Section III, we examine the effect of
adding an isotropic elastic scattering due to unitary scatterers
as well as the inelastic scattering due to spin fluctuations.
Finally,  section IV contains our comparison with existing data,
conclusions and plans for future work.

\section{Elastic Scattering}

\subsection{Normal State}

We first consider a model system where scattering occurs only
because of disorder.   For simplicity, we will assume that the
most important feature of the potential due to out-of-plane
impurities experienced by electrons moving in the CuO$_2$ planes
is its finite range $\kappa^{-1}$. We therefore model a single
impurity simply as a term $V(r)=V_0e^{-\kappa r}$, or
\begin{equation}
V_{\k\k'} = {2\pi \kappa V_0\over
\left((\k-\k')^2+\kappa^2\right)^{3/2} },\label{imppot}
\end{equation}
where $V_0$ sets the strength of the potential and $\kappa^{-1}$
is its range. Note that  $\k$ and $\k'$ are only defined up to a
reciprocal lattice vector.   The self-energy due to many such
impurities gives rise to an elastic broadening of quasiparticle
states which depends upon the position of ${\bf k}$ in the
Brillouin zone. For weak scattering, the Born approximation for
the retarded self-energy associated with a random distribution of
$n_I$ impurities per unit area has the usual form in the normal
state
\begin{equation}
{\Sigma}({\bf k},\omega) =  n_I \sum_{k^\prime} |V_{\k\k'}|^2
{G}^0(\k',\omega), \label{BornSigma}
\end{equation}
where $\omega$ is understood to include an infinitesimal positive
imaginary part, and the superscript on the single-particle Green's
function $G^0\equiv (\omega-\epsilon_\k+\mu)^{-1}$ indicates that
at first we ignore self-consistency, i.e. $G^0$ does not depend on
$\Sigma$.

\begin{figure}
\includegraphics[width=.9\columnwidth,clip,angle=0]{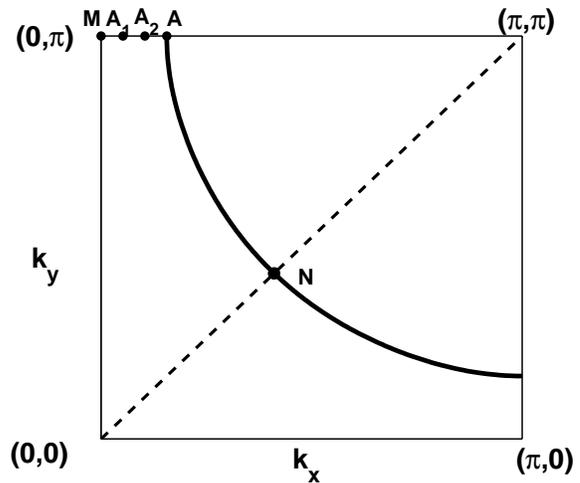}
\caption{The Fermi surface corresponding to the band $\epsilon_k$
given by Eq.~\eqref{three} with $t^\prime/t=-0.35$ and $\mu=-1$.
The energy distribution curves of $A(\k, \omega)$ will be
discussed for $\k=\k_N$ (nodal), $\k=\k_A$ (antinodal), and
various $k$ values along the $(0,\pi)$ to $\k_A$ cut shown by the
solid points.} \label{fig:FS}
\end{figure}

     To model the electronic structure we take a simple near-neighbor
hopping $t$ and
next-near-neighbor hopping $t^\prime$, such that
\begin{equation}
\epsilon_\k = - 2t (\cos k_x + \cos k_y) - 4t^\prime \cos k_x \cos
k_y - \mu \label{three}
\end{equation}
and set $t^\prime/t=-0.35$ and $\mu/t=-1$.  Note that with this choice
of parameters there is a van Hove singularity with a peak
in the total density of states at $-0.4t$.   The Fermi surface for
these parameters is shown in Fig.~\ref{fig:FS}; it is similar but
not identical to the Fermi surfaces found for both YBCO-123 and
BSCCO-2212 by ARPES.

\begin{figure}
\begin{center}
\includegraphics[width=.9\columnwidth,clip,angle=0]{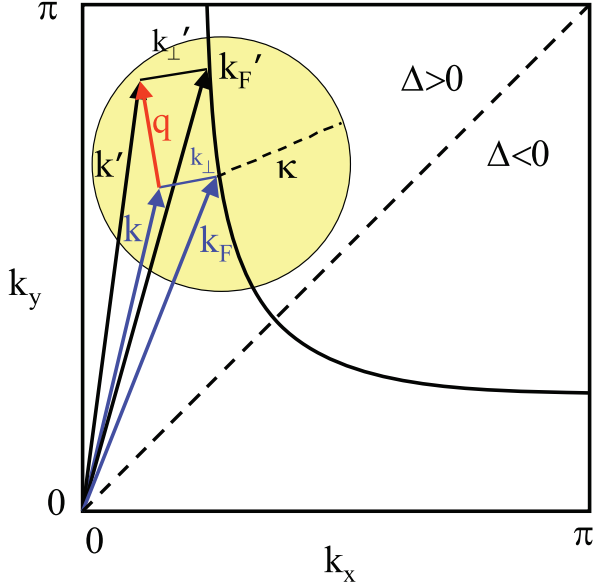}
\caption{Geometry for the forward scattering process in which a
quasiparticle scatters from {\bf k} to ${\bf k}^\prime$.}
\label{fig:scattrange}
\end{center}
\end{figure}

\begin{figure}
\includegraphics[width=\columnwidth,clip,angle=0]{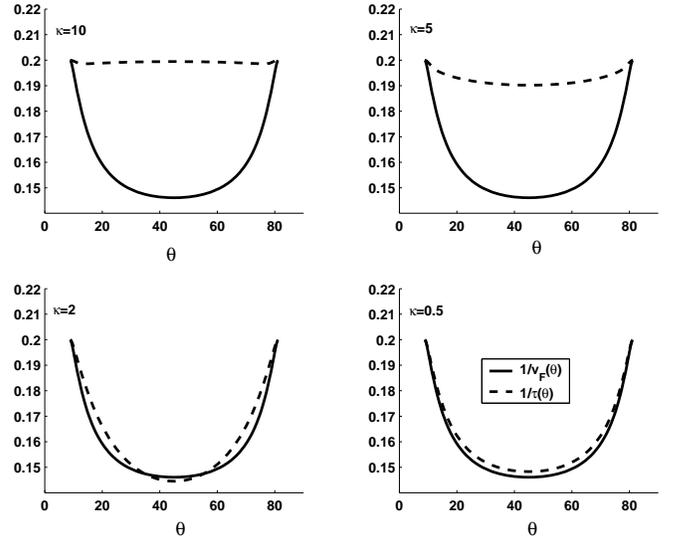}
\caption{The elastic scattering width $\Gamma({\bf k})= -{\rm
Im}\,\Sigma(\k_F, 0)$ in units of $t$ plotted around the Fermi
surface with $\theta=\tan^{-1} (k_y/k_x)$. Results are shown in
Fig.~2a--d for $\kappa=$ 10, 5, 2, and 0.5 respectively with
$n_I|V_0|^2$ adjusted so that $\Gamma_{max}=-{\rm
Im}~\Sigma(\k_A,0)$ is equal to $0.2t$. The dashed curve is
proportional to $v^{-1}_F(\theta)$ and one sees that as $\kappa$
decreases and the scattering peaks in the forward direction,
$\Gamma({\bf k})$ varies as $v^{-1}_F(\theta)$.} \label{fig:3}
\end{figure}

As the range of the potential $\kappa^{-1}$ increases, the
scattering of a quasiparticle from $\k$ to $\k'$ becomes peaked in
the forward direction.  As shown in Figure \ref{fig:scattrange},
when $\k$ is close to $\k'$ and both are not too far from the
Fermi surface, we may parameterize them as
\begin{eqnarray}
\k&=&\k_F+k_\perp\hat{k}_\perp\\
\k'&=&\k_F + \q_\parallel +k_\perp' \hat k_\perp'
\end{eqnarray}
where $\q=\k-\k'$ is the momentum transfer and $\q_\parallel$ its
component parallel to the Fermi surface.  The unit vectors $\hat
k_\perp$ and $\hat k_\perp'$ are the projections of $\k$ and $\k'$
onto the Fermi surface, respectively, such that, e.g.,
$\epsilon_{\k'}= v_F(\k')k_\perp'$.  The imaginary part of the
retarded self-energy (\ref{BornSigma}) becomes

 \begin{eqnarray}
\Sigma''(\k,\omega)&=& {-n_i (2\pi \kappa V_0)^2\over (2\pi)^2}
\int  {dk_\parallel' dk_\perp'\over \left[q^2
+\kappa^2\right]^{3}}~\delta(\omega-\epsilon_{\k'}) ~~~\\&\simeq&
-{n_i \kappa^2V_0^2\over |v_F(k_\parallel)|}\, \int
{dq_\parallel'~\over \left[q_\parallel'^2+  (k_\perp-{\omega \over
v_F})^2 +\kappa^2\right]^{3}
}~~~~\\
&\simeq& -\frac{3\pi n_i  V_0^2}{8 |v_F(k_\parallel)|\kappa^3}
\ \frac{1}{\Bigl[\Bigl(\frac{\epsilon_k -
\omega}{v_F(k_\parallel)\kappa}\Bigr)^2
+1\Bigr]^{\frac{5}{2}}}\, . ~~~\label{Sigapprox1}
\end{eqnarray}
Eq. (\ref{Sigapprox1}) shows explicitly that in the limit of small
$\kappa$, the self-energy  becomes more and more sharply peaked
``on the mass shell" $\omega=\epsilon_k$. This is a generic
feature of long-range potentials. For example, on the Fermi
surface $\k=\k_F$ at $ \omega=0$,
\begin{equation}
-\Sigma''(\k_F,0)\equiv \Gamma_0(\k_F) = {3 \pi n_i V_0^2\over 8
|v_F(\k_F)|\kappa^3},\label{Gamma0def}
\end{equation}
In Figure \ref{fig:3}, we show how this
angular dependence $\propto 1/|v_{\k_F}|$ is approached by the exact result
(\ref{BornSigma}) as $\kappa$ decreases and
 the range of the
scattering is increased. In the figure and in what follows in this
section,  we will use the result (\ref{Gamma0def}) with
$n_I|V_0|^2$ chosen such that $\Gamma_0(\k_A) = 0.2t$
corresponding to $\Gamma_0(\k_A)\simeq 30$ meV  for $t=.15$ eV.
From Fig.~3 we see that in the forward scattering limit,
$\Gamma_0(\k_N) \simeq \Gamma_0(\k_A)/1.4\simeq 0.14t$. We stress
that the precise dependence of the forward elastic part of the
self-energy (\ref{Gamma0def}) on $\kappa$ and on momentum depend
on the details of the Fermi surface shape and impurity scattering
potential.  However, we do not expect qualitative features of the
resulting spectra to be affected.   We note further that the
absolute magnitudes of the parameters chosen are roughly
consistent with the ${\cal O}$(10\%) weak scatterers of strength
$V_0$ of ${\cal O}(t)$ and range  of ${\cal O}$ (1-2$a$) extracted
from FT-STS data in Ref.~\cite{ZAH}. In Section III, we will show
results for various values of $\Gamma_0(\k_A)$ and $\kappa$.

\begin{figure}
\includegraphics[width=\columnwidth,angle=0]{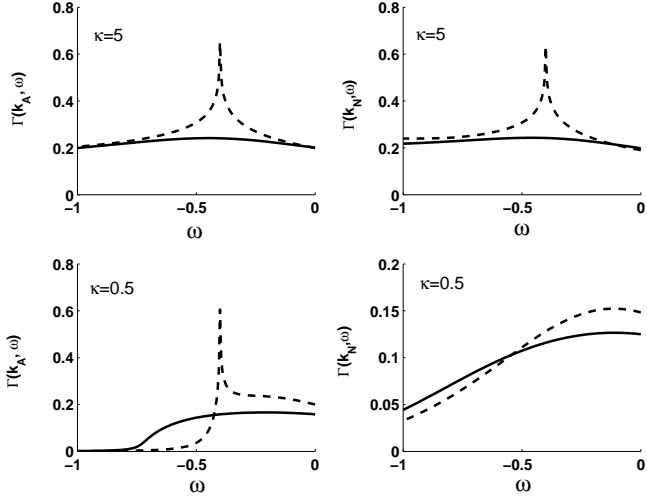}
 \caption{Scattering rate $\Gamma(\k,\omega)=- {\rm
Im}\,\Sigma(\k,\omega)$ for $\k$ on the Fermi surface at points A
(left) and N (right),
 for two different values of forward
scattering parameter $\kappa =$ 5 (top) and 0.5 (bottom). Here, as
in Fig.~\ref{fig:3},
 $n_iV_0^2$ has been chosen to give
$\Gamma(\k_A)=0.2t$. Dashed curves: non-self consistent Born
approximation; solid curves: self-consistent Born approximation.
}\label{fig:scattrate}
\end{figure}

Thus far we have not considered the effect of self-consistency,
i.e. replacing $G^0$ in Eq. (\ref{BornSigma}) by
$G(\k,\omega)=[(G^0)^{-1}-\Sigma]^{-1}$. In the normal metal one
is used to ignoring this distinction, as the self-consistent
solution for pointlike scatterers can be shown to be identical to
the non-self-consistent one up to corrections of order
$(\omega/E_F)^2$, where $E_F$ is the Fermi energy.  If the
scatterers have a finite range, however, this argument breaks down
and self-consistency becomes important. As seen in Figure
\ref{fig:scattrate}, the effect of self-consistency is to reduce
the frequency dependence of $\Sigma^{\prime\prime}(\k,\omega)$
induced by electronic structure; in particular, the van Hove
singularity at $\omega=-0.4t$ is eliminated.  This may account for
the complete absence of van Hove spectral features in STM and
other tunnelling experiments on BSCCO.  We note further that the
scattering rate is cut off in the forward scattering case
$\kappa=0.5$  when $|\omega|> v_F\kappa$, reflecting the fact that
if the electron's momentum can only be shifted by a small amount
$\sim \kappa$, the allowed energy transfer is also restricted.
\subsection{Superconducting State}

\subsubsection{Elastic Self-energy}

\begin{figure}
\includegraphics[width=\columnwidth,clip,angle=0]{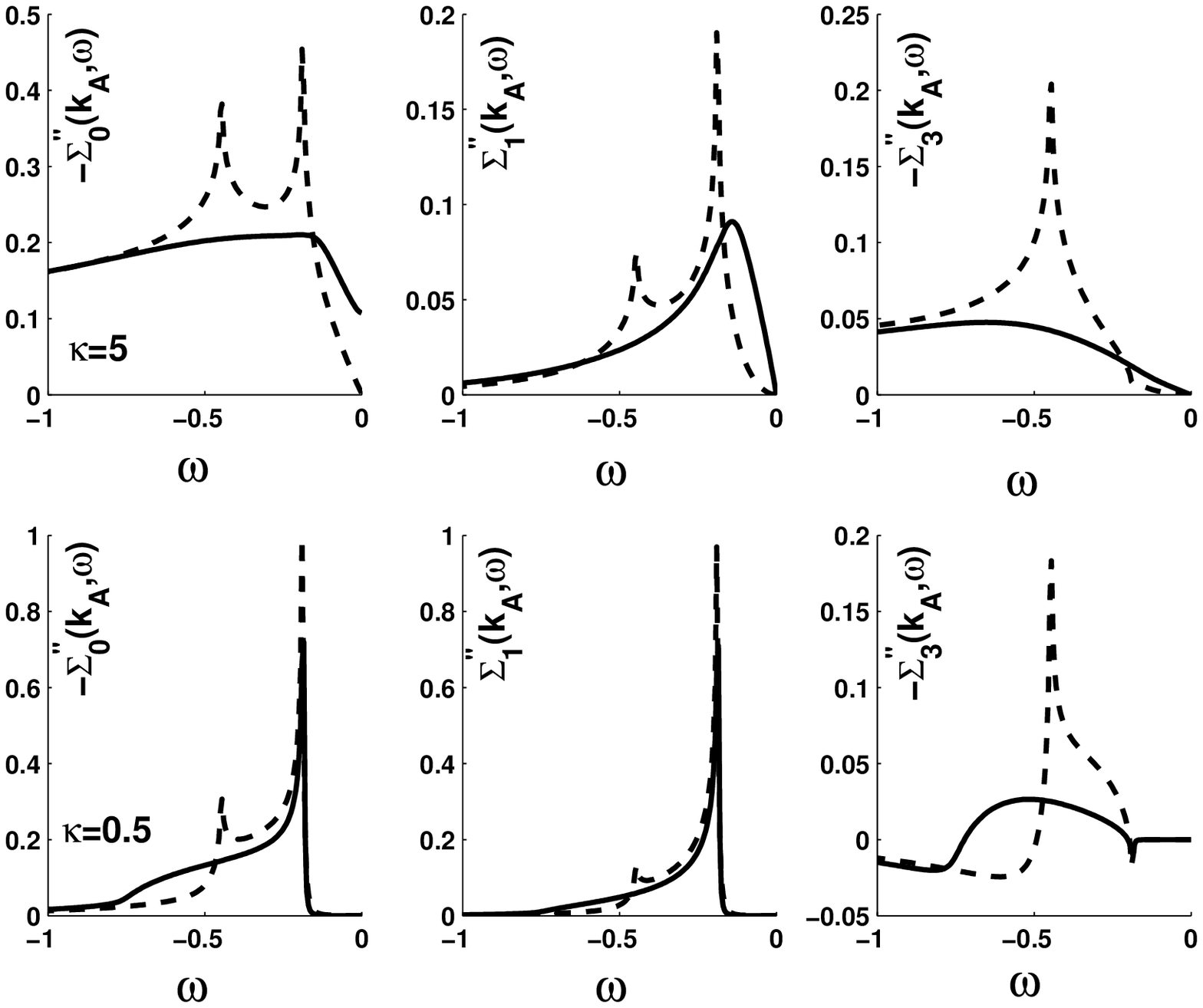}
\includegraphics[width=\columnwidth,clip,angle=0]{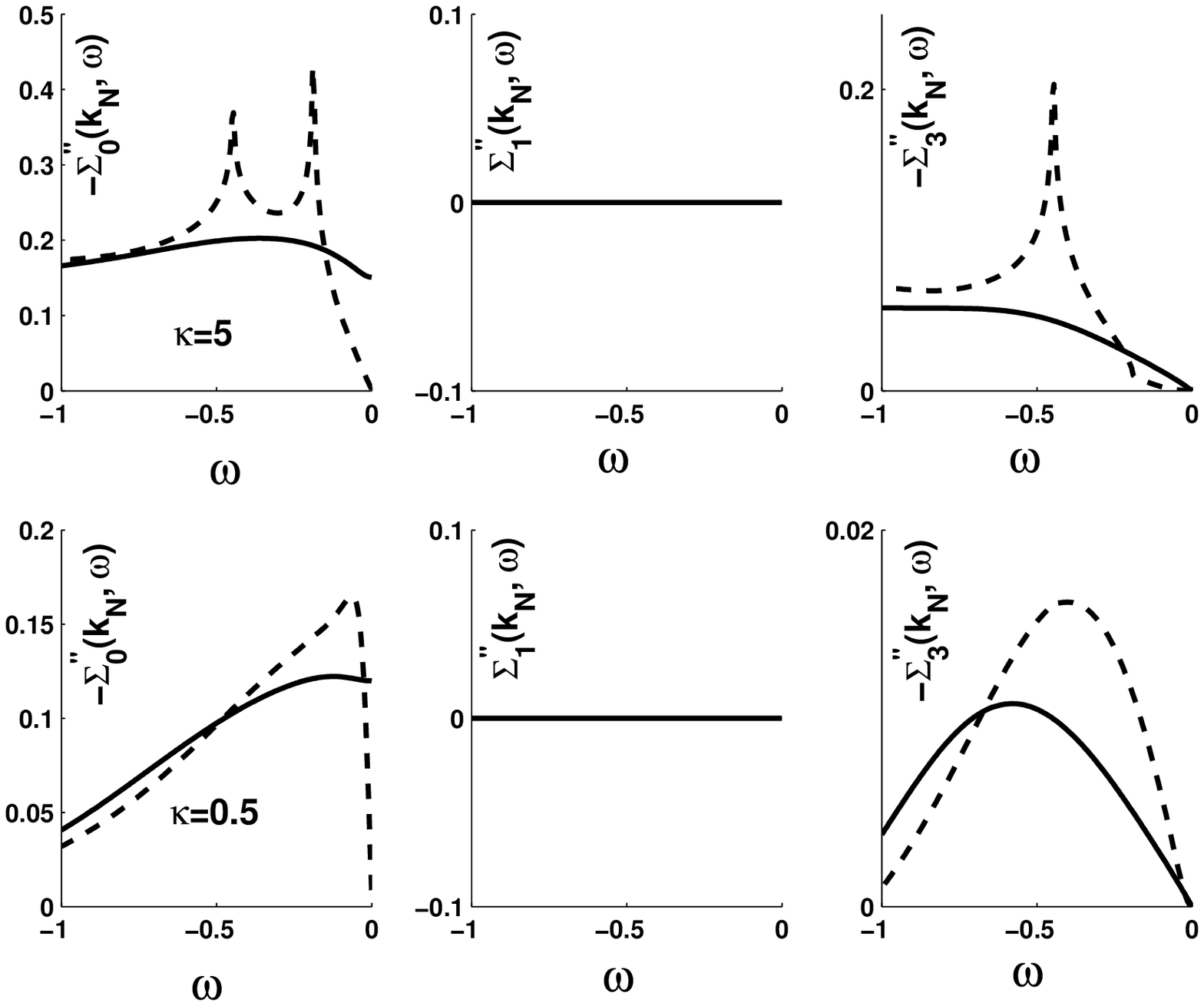}
 \caption{The self-energy terms -Im $\Sigma_0
(\k,\omega)$, Im $\Sigma_1 (\k,\omega)$, and -Im $\Sigma_3(\k,
\omega)$ in the superconducting state at $T=0$ for  $\k = \k_A$
(top) and $\k_N$ (bottom), for $\kappa=5$ and 0.5 and  the same
band and scattering parameters as previously used. Here
$\Delta_k=\Delta_0 \ (\cos_x-\cos k_y)/2$ with $\Delta_0=0.2t$.}
\label{fig:NambuSigs}
\end{figure}

\begin{figure}
\includegraphics[width=\columnwidth,clip,angle=0]{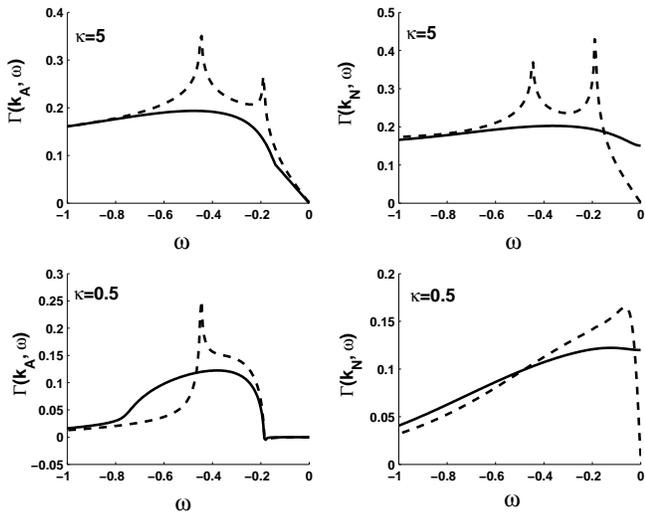}
 \caption{
 Scattering rate $\Gamma_{el}(\k,\omega)$  vs.~$\omega$ for $\k=\k_A$(left) and $\k_N$(right)
  in the superconducting state
at $T=0$, for
 $\kappa=5$ (top) and $\kappa=0.5$ (bottom).  Here
$\Gamma_0(\k_A)=0.2t$.  } \label{fig:effectiveSig}
\end{figure}

To describe the superconducting state, we assume a
BCS d-wave order parameter corresponding to pure nearest-neighbor
pairing,
\[
\Delta_\k = \frac{\Delta_0}{2}\ (\cos k_x - \cos k_y)
\]
with $\Delta_0/t=0.2$. The full matrix Green's function in the
presence of scattering in the superconducting state is
\begin{equation}
G(\k, \omega) = {\tilde\omega\tau_0
+\tilde\epsilon_\k\tau_3+\tilde\Delta_\k\tau_1\over
\tilde\omega^2-\tilde\epsilon_\k^2-\tilde\Delta_\k^2},
\label{fifteen}
\end{equation}
where $\tilde\omega\equiv \omega-\Sigma_0$,
$\tilde\epsilon_\k\equiv \epsilon_\k+\Sigma_3$, $\tilde \Delta_\k
\equiv \Delta_\k+\Sigma_1$, and the $\Sigma_\alpha$ are the
components of the self-energy proportional to the Pauli matrices
$\tau_\alpha$ in particle-hole space.  If we first assume the
simplest case, that the scattering is entirely elastic and weak,
we may approximate the self-energy in the Born approximation
similar to \eqref{BornSigma} as
\begin{equation}
\underline{\Sigma} = n_I \sum_{k^\prime} |V_{\k\k'}|^2\tau_3
\underline{G}^0(\k',\omega)\tau_3, \label{BornSC}
\end{equation}
with Nambu components
\begin{eqnarray}
\Sigma_0(\k,\omega) & = & {n_I}\,  \sum_{\k'}|V(\k, \k^\prime)|^2
\ \frac{\omega }{\omega^2-\epsilon^2_{\k^\prime} -
\Delta^2_{\k^\prime} }\, ,\label{six}\\
\Sigma_3(\k,\omega) & = & {n_I}\,  \sum_{\k'} |V(\k ,\k^\prime)|^2
\ \frac{\epsilon_{\k^\prime}}{\omega^2-\epsilon^2_{\k^\prime} -
\Delta^2_{\k^\prime} }\, , \label{seven}\\
\noalign{\hbox{and}} \Sigma_1(\k,\omega) & = &  -{n_I}\,
\sum_{\k'} |V(\k, \k^\prime)|^2 \
\frac{\Delta_{\k^\prime}}{\omega^2-\epsilon^2_{\k^\prime} -
\Delta^2_{\k^\prime} }\, . \label{eight}
\end{eqnarray}
We will calculate the  self-energies both non-self-consistently,
as in Eq. (\ref{BornSC}), and self-consistently by requiring that
$\underline{\Sigma}[G^{0}]\rightarrow \underline{\Sigma}[G]$. In
Figure \ref{fig:NambuSigs}, we show the variation of these
self-energy components with energy at $\k=\k_N$ and $\k=\k_A$.
Again the van Hove singularity is washed out in the
self-consistent evaluation, and it is furthermore noteworthy that
the $\Sigma_3$ component becomes quite small in the forward
scattering limit.
While in Figure \ref{fig:NambuSigs} gaps appear in the
$\Sigma_\alpha$ near the A point  for small $\kappa$, the values
of $\Sigma_0$ and $\Sigma_1$ near the gap edge $\omega \gtrsim
\Delta_\k$ are large, of order several times the hopping $t$.
Quasiparticle properties near the Fermi surface are determined,
however,  by particular combinations of the Nambu self-energy
components. As can immediately be seen from the denominator of
\eqref{fifteen}, the total elastic scattering rate broadening the
quasiparticle state of energy $\omega$ will be
\begin{equation}
\Gamma_{el} (\k) \simeq -{\rm Im} \left(\Sigma_0 (\k, \omega) +
\frac{\Delta_\k}{\omega}\ \Sigma_1 (\k, \omega)\right),
\label{totalscattrate}
\end{equation}provided one can neglect
$\Sigma_3$ (see Fig. \ref{fig:NambuSigs}).   In Figure
\ref{fig:effectiveSig}, we see that as $\omega$ goes to $-\Delta
(\k_A)$ for $\k=\k_A$,  $\Gamma_{el}$ is suppressed by the near
cancellation of the two components in (\ref{totalscattrate}) when
$\kappa$ becomes small.
 To obtain   some insight into the physical origins of this cancellation, we
derive approximate analytical forms  for $\kappa \ll 1$ following
the discussion of the normal state above, leading to

\begin{eqnarray}
\Sigma_\alpha''(\k,\omega)& \simeq&
 -
{\Gamma_0(\k_F)\over 2\sqrt{\omega^2 - \Delta_{\k}^2}}
\sum_{\nu=\pm 1} s_\alpha\times \nonumber\\
&& \times\left[\left( \frac{\epsilon_k-\nu
\sqrt{\omega^2-\Delta_\k^2}}{\kappa v_F(\k)}\right)^2
 +1\right]^{-\frac{5}{2}} .
\end{eqnarray}
Here $s_\alpha = |\omega|$, $-\Delta_\k \,{\rm sgn}\omega$, and
$\nu\, {\rm sgn} \omega \sqrt{\omega^2-\Delta_\k^2}$ for the Nambu
components $\alpha=0,1$ and 3 respectively, and $\Gamma_0 (\k)$ is
given by Eq.~\eqref{Gamma0def}.  Note that $\Sigma_3''$ vanishes
on the Fermi surface $\epsilon_\k=0$ in this limit. In order to
gain some further intuition for these expressions, we specialize
to the case where the momentum $\k$ is close to the Fermi surface
and the energies $\omega$ are small, such that $|(\epsilon_k \pm
\sqrt{\omega^2 - \Delta^2_k})/\kappa v_F(\k)|\ll 1$. The
self-energies may then be written
\begin{eqnarray}
\Sigma_0''(\k,\omega) & \sim & -\Gamma_0 (\k_F)
\ \frac{|\omega|} {\sqrt{\omega^2-\Delta^2_\k}} \label{Sig0forward}\\
\Sigma_1 ''(\k,\omega) & \sim & \Gamma_0 (\k_F)\
\frac{\Delta_\k  \,{\rm sgn}\, \omega}{\sqrt{\omega^2 - \Delta^2_\k}} \label{Sig1forward}\\
\Sigma_3''(\k,\omega)&\simeq & 0,\label{Sig3forward}
\end{eqnarray}
but are strongly supressed  due to energy conservation when
$|\sqrt{\omega^2-\Delta_\k^2}-\epsilon_\k|$ becomes greater than
$\kappa v_F$, as one may observe in Figure  \ref{fig:NambuSigs}.
These equations  are now identical in form to
 those expected for an $s$-wave
 superconductor (even when self-consistency is included).
  All peculiarities of the $d$-wave state which
result from momentum averaging over the Fermi surface have
disappeared. We therefore expect {\it a priori} to recover
Anderson's theorem, the insensitivity of bulk thermodynamic
properties to nonmagnetic scattering \cite{Andersontheorem}. The
physical reason for this is clear: as seen in Figure
\ref{fig:scattrange}, for small $\kappa$ and $\k$ a distance at
least $\kappa$ from the node, small angle scattering cannot mix
order parameters of different signs, and therefore cannot break
Cooper pairs. The analogy with the $s$-wave superconductor and
relation to Anderson's theorem in the context of ARPES is
discussed further in Appendix 1. In addition, the slowing of the
rate of $T_c$ suppression due to disorder as scattering becomes
more anisotropic was  treated some time ago by several authors
\cite{Tcsupress}. These works were motivated by $T_c$ suppression
rates in the cuprates which appear to be 2-3 times slower than
predicted by the classic Abrikosov-Gorkov formula \cite{AG}
appropriate for for pointlike isotropic scatterers. The
suppression of $T_c$ near the pure forward scattering limit has
been discussed recently in detail by H.-Y. Kee \cite{Kee01}.

  In the forward scattering limit where (\ref{Sig0forward}-\ref{Sig3forward})
  hold, the  effective elastic scattering rate (\ref{totalscattrate}) becomes
\begin{equation}
\Gamma_{el} (\k,\omega) \simeq \Gamma_0 (\k_F)
\frac{\sqrt{\omega^2-\Delta^2_\k}}{|\omega |} \ \quad , \quad
|\omega|\gtwid |\Delta_\k|\, . \label{totalscattrateA}
\end{equation}
For $\k$ along the nodal direction, the elastic broadening in the
superconducting d-wave state is equal to its value in the normal
state. However, for $\k$ in the antinodal region, the broadening
vanishes as $\omega \to \Delta_\k$ and approaches the normal state
value only when $\omega$ becomes large compared with $\Delta_\k$.
The elastic contribution $\Gamma_{el}(\k_A, \omega)$ to the
broadening at the antinodal $\k_A$ point versus $\omega$ is shown
in Fig.~\ref{fig:effectiveSig}. Physically, the individual
contributions to the normal $\Sigma_0 (\k, \omega)$ and anomalous
$\Sigma_1 (\k, \omega)$ self-energies are both enhanced by the
density of states factor $(\omega^2 - \Delta^2
(\k))^{-\frac{1}{2}}$ (Fig. \ref{fig:NambuSigs}). However, the
normal contribution describing the scattering out of state $\k$
into $\k^\prime$ is compensated by the anomalous contribution
scattering into $\k$ from the pair condensate.
This gives rise to the
suppression of the elastic scattering seen in
Fig.~\ref{fig:effectiveSig} relative to Fig. \ref{fig:NambuSigs}
as $\omega$ approaches $-|\Delta_{\k_A}|$ from below.

\subsubsection{Spectral Function}

The near-cancellation of the two Nambu components of the
self-energy near the gap edge in the forward scattering limit
leads to a dramatically reduced {\it elastic} scattering rate in
the superconducting state, which
 sharpens the spectral features of quasiparticles which are not too
close to the nodes.
 Within the current model where we continue to neglect
inelastic scattering due to electron-electron interactions, we now
turn to the 1-electron spectral function measured by ARPES. In the
forward scattering limit, with the self-energy given by Eqs.
\eqref{Sig0forward}-\eqref{Sig3forward}, one obtains a result for
the Green's function previously discussed by
Markiewicz\cite{Mar03},
\begin{equation}
\underline G(\k, \omega) \simeq \frac{(\omega \tau_0
+\Delta_\k\tau_1)z(\k, \omega) + \epsilon_k\tau_3  }
 {(\omega^2-\Delta_\k^2)z(\k, \omega)^2-\epsilon^2_k} \, .
\label{eighteen}
\end{equation}
Here $z(\k,\omega)=1+ i\Gamma_0(\k) {\rm sgn}\,\omega
/\sqrt{\omega^2-\Delta^2_k}$. The electron component of the
spectral function is then
\begin{eqnarray}
A(\k,\omega)&=&-{1\over \pi}  {\rm Im}
\,G_{11}(\k,\omega)\nonumber\\
&=& -{1\over \pi}\,  {\rm Im} \frac {\omega z(\k,\omega)
+\epsilon_\k}{(\omega^2-\Delta_\k^2)z(\k,\omega)^2-\epsilon^2_k}
\, .\label{Akw}
\end{eqnarray}
It is useful to consider a few special cases of \eqref{Akw} more
closely. In particular, on the Fermi surface $\epsilon_\k=0$ one
has the simple expression
\begin{equation}
A(\k_F,\omega) = {\Gamma_0(\k)\over \pi}{1\over
\sqrt{\omega^2-\Delta_\k^2}}\, {|\omega|\over \omega^2-\Delta_\k^2
+\Gamma_0(\k)^2},\label{spectral1}
\end{equation}
while near the gap edge, e.g. $\omega \lesssim -|\Delta_\k|$,
\begin{equation}
A(\k,\omega) \simeq \frac{1}{\pi}\
 \ \frac{\Gamma_0 (\k)}{\epsilon^2_k +
\Gamma^2_0 (\k)} \ \frac{|\omega|}{\sqrt{\omega^2-\Delta_\k^2}}\,
. \label{spectral2}
\end{equation}

At the nodal point $\k_N$, where the gap vanishes, the spectral
weight is given by the simple Lorentzian form
\begin{equation}
A(\k_N,\omega) =  {\Gamma_0(\k_N)/\pi\over \omega^2
+\Gamma_0(\k_N)^2}, \label{twentyfive}
\end{equation}
and at low temperatures where the elastic scattering is dominant
one can determine $\Gamma_0(\k_N)$. However, for $\k$ at the
antinodal point $\k_A$ such that  $-\Delta_{\k_A} - \delta \omega
< \omega < -\Delta_{\k_A}$,
\begin{equation}
A(\k_A, \omega) \cong \frac{1}{\pi}\
\frac{\Delta_{\k_A}}{\Gamma_0(\k_A)} \
\frac{1}{\sqrt{\omega^2-\Delta^2_{\k_A}}}\, , \label{twentysix}
\end{equation}
where the ``width" $\delta \omega$ depends upon the ratio
$\Delta_{\k_A}/\Gamma_0(\k_A)$.  If $\Delta_{\k_A}$ is large
compared with $\Gamma_0(\k_A)$, the width $\delta\omega \simeq
\Gamma_0(\k_A)^2/(2\Delta_{\k_A})$.  If one integrates
$A(\k_A,\omega)$ from $-\Delta_{\k_A}-\delta\omega$ to
$-\Delta_{\k_A}$ to define a ``peak intensity"

\begin{eqnarray}
I(\k_A)=\int_{-\Delta_{\k_A} - \delta \omega}^{-\Delta_{\k_A}}
A(\k_A,\omega)\,d\omega \simeq {1\over \pi},
\end{eqnarray}
which is independent of $\Delta_{\k_A}/\Gamma_0(\k_A)$.  However,
when the system is sufficiently dirty such that
$\Delta_{\k_A}<\Gamma_0(\k_A)$, the falloff of $A(\k_A,\omega)$ as
$\omega$ decreases below $-\Delta_{\k_A}$ varies as
$(\omega^2-\Delta_{\k_A}^2)^{-1/2}$.  In this case, the scale is
set by $\Delta_{\k_A}$ and if one takes $\delta\omega =
\Delta_{\k_A}$, the peak intensity varies as $I(\k_A)\sim
\Delta_{\k_A}/\Gamma_0(\k_A)$. This is quite different from the
usual BCS quasiparticle result which is proportional to the
quasiparticle renormalization factor $z(\k_A)$ times a coherence
factor which is 1/2 on the fermi surface. It should be possible to
test the foward scattering scenerio by comparing the variation of
$I(\k_A)$ with $\Delta_{\k_A}/\Gamma_0(\k_A)$.

\begin{figure}
\includegraphics[width=\columnwidth,clip,angle=0]{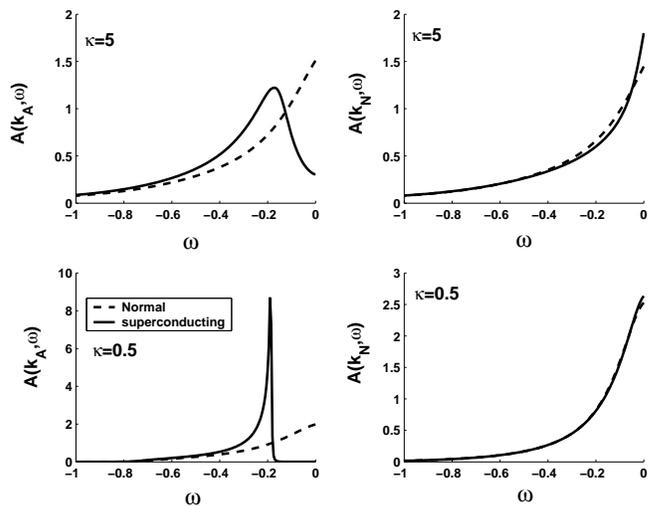}
\caption{Comparison of the $T=0$ self-consistent normal state
(dashed) and superconducting (solid) one-electron spectral
functions for $\kappa=5$ (top) and 0.5 (bottom) at $\k=\k_A$
(left) and $\k_N$ (right), with $\Gamma_0(\k_A)=0.2t$.}
\label{fig:Akwnormal}
\end{figure}

Numerical results for the normal and superconducting spectral
weights for $\omega <0$ using the self-consistent version of the
self-energy (\ref{six}-\ref{eight}) with $\omega\rightarrow
\tilde\omega$, etc.,  for the model impurity potential (1) are
shown in Fig.~\ref{fig:Akwnormal} for $\kappa =5$ and 0.5. For
$\kappa=0.5$, the scattering is predominantly forward and in the
superconducting state at $T=0$, the spectral weight for $\k_A$ is
seen to sharpen at $-\Delta_{\k_A}$ despite the fact that the
normal state broadening $\Gamma(\k_A)$ is of order the peak
position. For $\k=\k_N$, the nodal spectral weight has the
expected Lorentzian form with a width set by $\Gamma(\k_N)$. For
$\kappa=5$, corresponding to a more isotropic scattering, the
spectral weight for $\k=\k_A$ broadens. In the limit of isotropic
impurity scattering, the $\Sigma_1$ component of the impurity self
energy vanishes and $A(\k_A, \omega)$ has a Lorentzian-like
broadened peak.

\section{Isotropic Elastic and Inelastic
 Electron-Electron Scattering}

 In addition to the forward scattering by out-of-plane impurities,
in this section we consider  unitary limit isotropic elastic
scatterers as well as inelastic electron-electron collisions.  In
fact, it is the momentum and frequency dependence of the latter
interaction about which one hopes to learn more
 from the ARPES spectrum. Here we proceed
phenomenologically by writing

\begin{equation}
\underline{\Sigma}_{tot}=
\underline{\Sigma}_{el,f}+\underline{\Sigma}_{\rm el,u} +
\underline{\Sigma}_{\rm inel}.
\end{equation}
The first term is the contribution from elastic (quasi-) forward
scatterers we have discussed in Section II.  The second term represents
the effect of   unitary scatterers (possibly Cu vacancies) with
concentration roughly $n_u\sim 0.2\%$ observed as zero-bias
resonances in STM experiments. It will be treated as usual in the
self-consistent $T$-matrix approximation,
\begin{equation}
\underline{\Sigma}_{\rm el,u} = - {n_u\over \sum_\k
G(\k,\omega)}\tau_0.
\end{equation}
In the normal state we find a scattering rate of $\Gamma_u \simeq
10^{-3}t$,  leading to an impurity bandwidth $\gamma_u \sim
\sqrt{\Gamma_u \Delta_0} \sim 10^{-2}t$ (of order 1 to 2 meV). We
note that the {\it width} of the resonance observed in STM is in
fact roughly 3 times this number, consistent with the
self-consistent T-matrix calculation\cite{AZH03}; impurity
resonances are visible up to about 10 meV in experiments.
Nevertheless the isotropic part of the elastic scattering appears
to have a relatively insignificant effect on the ARPES spectral
function, as shown below.
\begin{figure}
\includegraphics[width=\columnwidth,clip,angle=0]{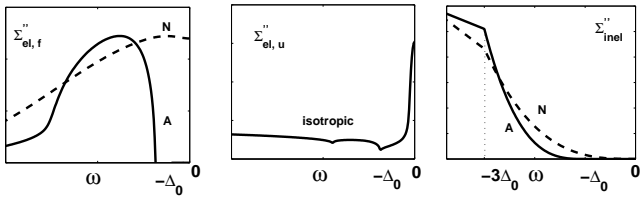}
\caption{Schematic depiction of the various contributions to the
scattering rate in the superconducting state at $T=0$.  First
panel: elastic forward scattering for $\k=\k_A$ and $\k_N$; Second
panel: isotropic unitarity limit impurity scattering due to $\sim
0.2\%$ impurities; Third panel: spin fluctuation inelastic
scattering rate interpolated from Ref. \cite{QHS96}}
\label{fig:cartoon}
\end{figure}

For the imaginary parts of the inelastic self-energies, we will
use numerical results obtained from a spin-fluctuation calculation
of the quasiparticle scattering. Following the argument leading
to (\ref{totalscattrate}), we define an effective {\it inelastic}
scattering rate
\begin{equation}
\Gamma_{\rm inel}\, (\k, \omega) = -{\rm Im} \left(
\Sigma_{0}^{\rm inel}(\k,\omega) + {\Delta_\k\over \omega}
\Sigma_{1}^{\rm inel}(\k,\omega) \right).\label{thirteen}
\end{equation}
In spin-fluctuation calculations of $\Gamma_{\rm inel}\, (\k,
\omega)$, it was found \cite{Quinlan94,Yashenkin} that at the
nodal point, at low temperatures, the scattering rate initially
increased as the third power of $\omega$ or $T$ depending upon
which is larger. At other $\k$ points on the Fermi surface, the
scattering rate varies approximately as the third power of this
energy measured relative to $\Delta_\k$. The reduction of the
inelastic scattering rate at low excitation energies reflects the
suppression of the low energy spin-fluctuations due to the opening
of the d-wave gap. Here, we will use a numerical interpolation of
the $\omega$-and $T$-dependent $\Gamma_{\rm inel}$ obtained from
Ref. \cite{QHS96}.

\subsection{The Antinodal Spectrum}

The various contributions to the self-energy are sketched in the
schematic diagram shown in Fig.~\ref{fig:cartoon}. We will use the
parameters discussed above to set the magnitudes of $\Sigma_{\rm
inel}$ and $\Sigma_{\rm el, u}$. Then we will consider various
values of $\kappa$ and $\Gamma_0(\k_A)$, characterizing the
forward elastic scattering.  While the effect of $\Sigma_{\rm
inel}$ is already contained in the intrinsic $T_c$ and the
suppression of $T_c$ due to $\Sigma_{\rm el, u}$ is negligible for
the parameters considered, this is not necessarily the case for
the forward elastic scattering. As discussed by Kee\cite{Kee01},
the suppression of $T_c$ varies as $\kappa^3$, so that for small
values of $\kappa$ such as $\kappa=0.5$, the suppression of $T_c$
is negligible for the scattering rates $\Gamma_0(\k_A)$ which we
will consider. However, at larger values of $\kappa$ this is not
the case, so that the $T_c$ shift or the lack of shift implies
constraints on $\kappa$ and $\Gamma_0$. For the moment, however,
we will ignore this and simply compare $\kappa=0.5$ with
$\kappa=2$ for three different scattering rates
$\Gamma_0(\k_A)=0.2$, 0.1, and 0.05 in units of $t$ corresponding
to $\Gamma_0(\k_A)/\Delta_0=1$, 0.5, and .025.

\begin{figure}
\includegraphics[width=.6\columnwidth,clip,angle=0]{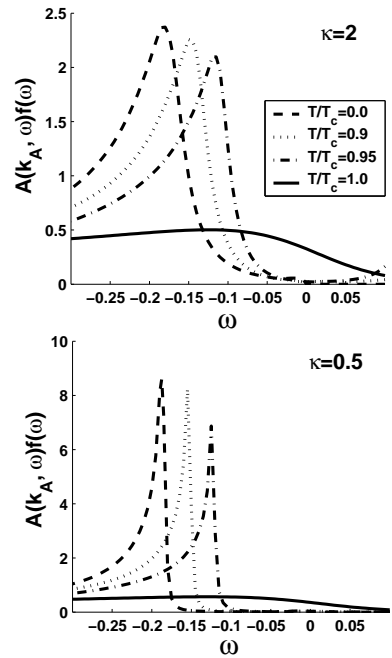}
\caption{Finite temperature spectral function at the antinodal
point A on the Fermi surface multiplied by the fermi function,
$A(\k_A,\omega)f(\omega)$ vs.~$\omega$ including the full model
self-energy as described in Section III. Results for $\kappa=2$
and 0.5 with $\Gamma_0(\k_A)=0.2t$ are shown.} \label{fig:EDC_A}
\end{figure}

In Fig.~\ref{fig:EDC_A} we show the temperature dependence of the
energy distribution curves $A(\k, \omega)\, f(\omega)$ for $\k$ at
the antinodal point $\k_A$. Here $f(\omega)$ is the Fermi
function. In the superconducting state there is a square-root-like
behavior as $\omega$ approaches $-|\Delta_{\k_A}|$ from below.
This should be contrasted with the broad Lorentizian peak in the
normal state which is cut off by $f(\omega)$. While one could
understand that inelastic broadening would diminish as the
temperature is lowered, the asymmetric, one-sided,
square-root-like sharpening of the spectrum in the superconducting
state is a consequence of forward elastic scattering as discussed
in Section II. The peak intensity at $\k_A$ should scale as
$\Delta_0(T)$.

\subsection{Quasiparticle Dispersion Near the Antinodal Point}

\begin{figure}
\leavevmode\includegraphics[width=1.\columnwidth]{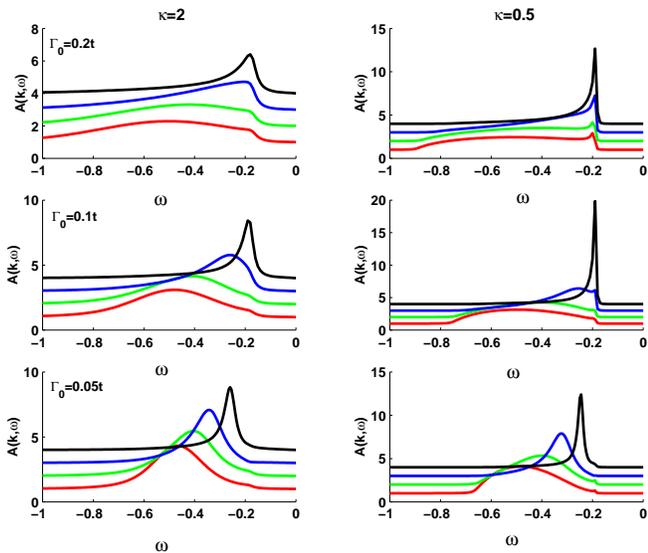}
 \caption{$A(\k,\omega)$ vs.~$\omega$ for $\kappa=2$ and 0.5.
Results are given for the $k$ points as indicated along the
 $(0,\pi)$- A cut in Fig.~\ref{fig:FS}. The disorder levels correspond to
$\Gamma_0(\k_A)/\Delta_0=1$, 0.5, and 0.025.  Note the spectra for
different $\k$ points have been offset for clarity.}
 \label{dispersionA}
\end{figure}

In a clean  superconductor, there is a peak in the spectral
function $A(\k,\omega)$ at the quasiparticle pole
$\omega=E_\k\equiv \sqrt{\epsilon_\k^2+\Delta_\k^2}$. In
particular, as the momentum moves along the cut from $(\k_A, \pi)$
to $(0, \pi)$ shown in Fig.~1, one expects to see a dispersion of
this peak to higher energies.
However, if the forward elastic scattering strength
$\Gamma_0(\k_A)\gtrsim \Delta_{\k_A}$, then  the peak in
$A(\k,\omega)$ remains at $-\Delta_\k$ rather than dispersing.
Figure \ref{dispersionA} shows plots of $A(\k,\omega)$ for
different values of $\k$ between the M and A points for $\kappa=2$
and 0.5, and several values of the scattering rate
$\Gamma_0(\k_A)$.

As samples improve, there is a natural tendency in this
 model for the spectrum for $\k$ not too far from the antinode
 to cross over from
one characterized by a nondispersive peak at $\Delta_\k$ in the
dirty limit where $\Gamma_0\sim \Delta_0$ to one characterized by
a dispersive quasiparticle peak at $E_\k$ when $\Gamma_0$ is small
compared to $\Delta_0$. This crossover is due to the way in which
the forward elastic scattering rate for a d-wave superconductor is
reduced as the gap edge is approached and is analogous to the same
effect discussed analytically in Appendix 1 for an $s$-wave
superconductor.    In a system with $\Gamma_0\gtrsim \Delta_0$, no
quasiparticle peak is observed,  but a sharp feature {\it does}
appear at $-\Delta_\k$,  representing simply the spectral weight
in the overdamped quasiparticle peak piling up at the gap edge as
in the $s$-wave case.
Only when $\Gamma_0$ becomes small compared to $\Delta_0$ does one
see a true quasiparticle peak dispersing as $-E_k$.  In the most
strongly forward scattering case, $\kappa=0.5$, one can see that,
depending on the strength of the scattering rate, one can have
simultaneously a broadened dispersing feature as well as a gap
edge feature.  It is tempting to speculate that this phenomenon is
related to the peak-dip-hump features observed generically below
$T_c$ in cuprate ARPES experiments, but we have not yet explored
this issue in detail.

\subsection{The Nodal Spectrum}
\begin{figure}
\includegraphics[width=\columnwidth,clip,angle=0]{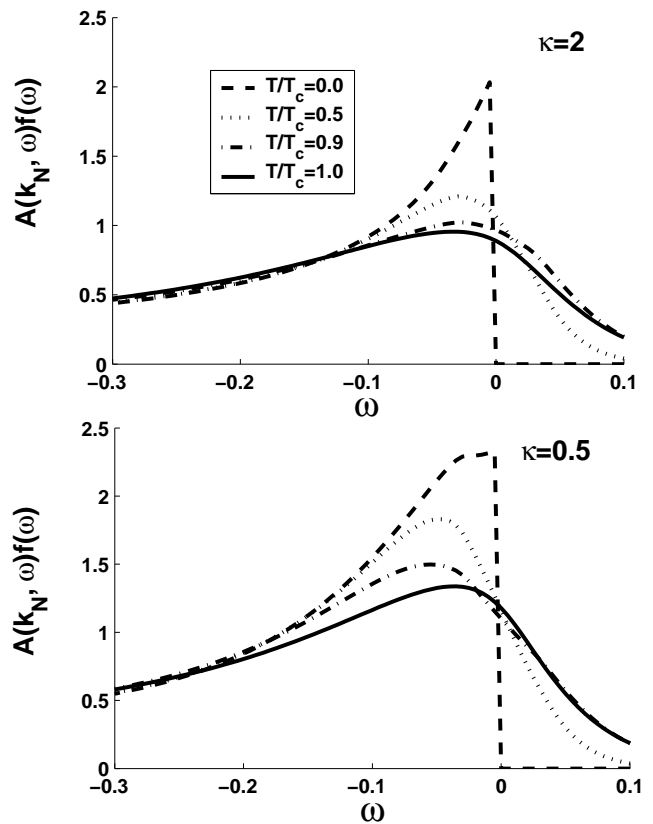}
\caption{Finite temperature spectral functions at the nodal point
N on the Fermi surface,  $A(\k_N,\omega)f(\omega)$ vs.~$\omega$
including the full model self-energy as described in Section III
for $\kappa=2$ and 0.5, with $\Gamma_0(\k_N)=0.14t$.}
\label{fig:EDC_C}
\end{figure}
In Figure \ref{fig:EDC_C}, the corresponding EDC's for the nodal
$k_N$ point are shown.  In this case, the $T=0$ spectral function
is a Lorentzian centered at the Fermi level, whose width is
limited essentially by the elastic scattering. For  the band
structure parameters we have chosen, $\Gamma_0(\k_N) \cong
\Gamma_0 (\k_A)/1.4$, so that $\Gamma_0(\k_N)=0.14t$. Results in
Fig.~\ref{fig:EDC_C} are shown for several different values of
$\Gamma_0(\k_N)$. At finite temperatures the peak is further
broadened by inelastic processes. Defining a width of the
asymmetric EDC's is quite difficult, a natural result of the
strongly $\omega$-dependent self energy in the current
approximation.  In fact, even the shape of the EDC curve is
difficult to compare directly to experiment, since it can be
qualitatively changed by a small amount of averaging along the
Fermi surface due to the angular resolution of the detector.  If
one goes a short angular distance away from the node along the
Fermi surface, one finds at low $T$ not a smooth Lorentzian for
$\omega<0$, but rather a square root singularity with a small
local gap $\Delta_\k$.  Averaging over a small $k$ region will
therefore make the ``leading edge" of this spectrum appear much
sharper.
\begin{figure}
\includegraphics[width=\columnwidth,clip,angle=0]{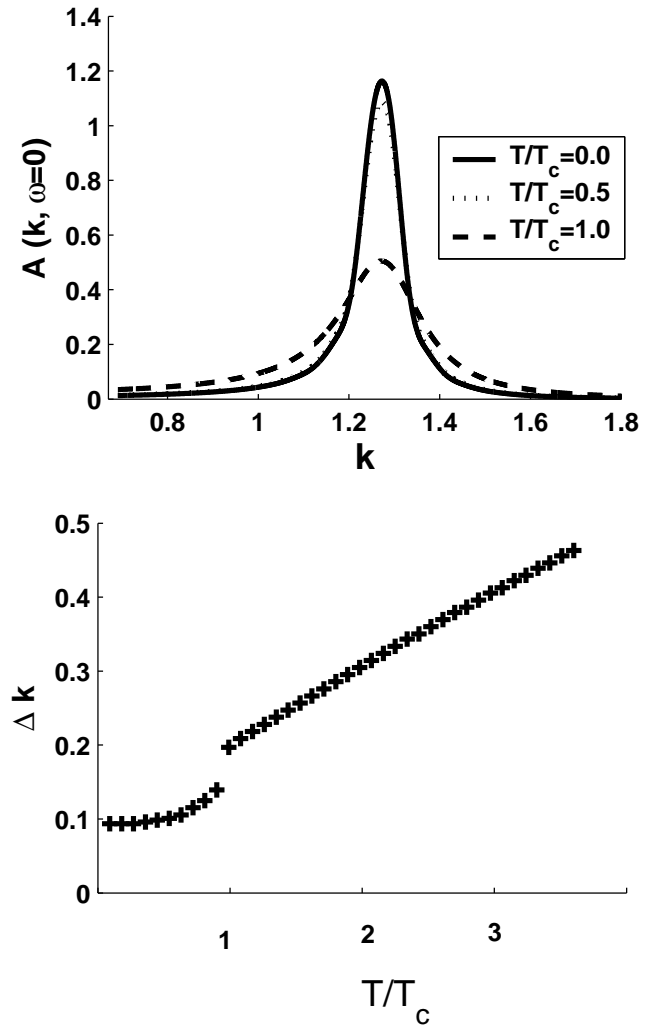}
\caption{a) Finite temperature spectral function near the nodal
point N along a cut perpendicular to the Fermi surface at N,
$A(\k,\omega=0)$ vs. $\k$, including full model self-energy with
$\kappa=0.5$ and $\Gamma_0(\k_N) = 0.14t $. b) The half-width of
$A(\k,\omega=0)$ at half-maximum, $\Delta k$, plotted
vs.~$T/T_c$.} \label{fig:MDC_C}
\end{figure}

 The intrinsic $T$-dependent broadening is therefore most clearly seen in the
momentum distribution curves (MDC's), which in this work (see
Eq.~\eqref{twentyfive}) are simply Lorentzians centered at the
Fermi surface.   In Figure \ref{fig:MDC_C}, we show that, within
this model, the rate of sharpening of the nodal MDC's indeed
increases somewhat in the superconducting state as the temperature
is lowered, but that the widths saturate at a value determined by
the elastic scattering.  The temperature dependence of $\Delta k$
shown in Fig. \ref{fig:MDC_C} is in disagreement with the results
of Valla et al.\cite{Val00}, who reported a linear MDC width
dependence on temperature even in the superconducting state.
However, the type of $\omega$ and $T$ dependence we find would
appear to be consistent with nodal EDC linewidth measurements of
Kaminski et al \cite{Kam00}.

\section{Conclusions}

The main point of this work is to raise the possibility that some
of the ARPES observations on optimally or overdoped samples which
appeared to be in conflict with the BCS results may in fact fit
within the BCS framework when the effects of forward elastic
scattering are taken into account. Thus, while forward elastic
scattering can be responsible for some of the anomalous width of
the spectral function measured above $T_c$, this need not be in
conflict with the observation of a sharp spectral feature in the
antinodal region below $T_c$.     Furthermore, if $\Gamma_0(\k_A)$
is larger than $\Delta(\k_A)$, the intensity associated with the
area under this feature varies as $\Delta(\k_A)/\Gamma(\k_A)$,
which depends upon the doping and temperature. At the same time,
the spectrum at the nodal point can exhibit a Lorentzian behavior
with a width that evolves smoothly through $T_c$ and then
partially narrows as the inelastic scattering is suppressed by the
opening of the gap. We have seen that these effects occur because
of a phenomenon similar to Anderson's theorem which applies for
much of the Fermi surface of a $d$-wave superconductor in the
forward scattering limit: if scattering is sufficiently peaked in
the forward direction, the scattering process does not mix states
with different signs of the order parameter, and no pairbreaking
occurs.  Sufficiently near the nodal point, however, the sign
change always takes place, implying that the full width of the
quasiparticle due to elastic scattering is recovered at the nodal
$\k_N$.

We have discussed simple approximate forms as well as fully
self-consistent numerical calculations  for the elastic
self-energies.  In the overall model for the self-energy, we also
included an approximate treatment of electron-electron collisions,
as well as unitary ``native defect" scatterers observed to be
present in the BSCCO-2212 material in STM experiments.  The energy
distribution curves found in the self-consistent calculations
using this model show no influence of the van Hove singularity,
and display an asymmetric shape with a rounded square root peak
near the local gap edge. Widths and temperature dependences can be
obtained which appear comparable to experiment, but it is
difficult to compare directly because of the unknown ARPES
background signal and bilayer splitting, which was not included
here. We also discussed the temperature dependence of the momentum
distribution curves in the superconducting state at the nodal
point.  It was found to have a Lorentzian shape which narrows with
decreasing temperature in the superconducting state, due to the
suppression of inelastic scattering, saturating at a value
determined by the elastic scattering.

Along the momentum cut $(\k_A, \pi)$ to $(0, \pi)$, if
$\Gamma_0(\k_A)$ is small compared to $\Delta_\k$, we find a
dispersing quasiparticle peak at $\omega=-E_k$. However, when
$\Gamma_0(\k_A)$ is comparable with $\Delta_\k$, the maximum
response occurs for $\omega \simeq -\Delta_\k$, does not disperse
as $\k$ moves away from the Fermi level, and can still represent a
sharp spectral feature.
This is similar to some of the older data on the BSCCO-2212
system, which has not been explained at this writing and might
have been taken for  some new type of dispersionless excitation
 in the superconducting state. However, as we have shown,
it is simply the consequence of forward elastic scattering which
is suppressed at the gap edge in the superconducting state.

Here we have focussed primarily on the effects of forward elastic
scattering on the BSCCO-2212 ARPES spectrum, and found that a
model of several per cent weak out-of-plane scatterers with a
range of order one lattice spacing, similar to current models of
Fourier transform STM measurements, can explain many qualitative
features of the  ARPES data. We believe that the unique way in
which this material is doped and disordered using current crystal
growth techniques endows it with an effective disorder potential
which strongly influences the low-energy quasiparticle properties
in a way which is characteristically different from the much
cleaner YBCO system, for example.  This picture should have
important and calculable consequences for other superconducting
properties, such as microwave conductivity, which we explore
elsewhere.

\acknowledgments

The authors thank E. Abrahams, N. Ingle,  D. Maslov, M. R. Norman,
and Z.X. Shen for enlightening discussions.  Partial support was
provided by ONR N00014-04-0060 and NSF DMR02-11166.

\appendix\section{One-electron spectral function in a disordered
$S$-wave superconductor}

Insight into the spectral function of a $d$-wave superconductor
with $\k$ near the antinodal point and reasonably forward
scattering can be gained by examining the spectral function of an
$s$-wave superconductor with disorder.  Although quite simple, we
are not aware that the spectral weight for an s-wave
superconductor with isotropic impurity scattering has been
discussed elsewhere.   For simplicity, we restrict our
consideration to isotropic, weak, nonmagnetic scatterers. The
self-energies  in the Born limit for a system with particle-hole
symmetry and an isotropic order parameter $\Delta$ are

\begin{eqnarray}
\Sigma_0&=& n_i V^2 \sum_\k {\tilde\omega\over
\tilde\omega^2-\epsilon_\k^2-\tilde\Delta^2}=
-i\Gamma {u\,{\rm sgn}\,u'\over \sqrt{u^2-1}}\\
\Sigma_1&=& n_i V^2 \sum_\k
{\tilde\Delta\over \tilde\omega^2-\epsilon_\k^2-\tilde\Delta^2}=i\Gamma
{\,{\rm sgn}\,u'\over \sqrt{u^2-1}} \\
\Sigma_3&=&0,
\end{eqnarray}
where $u=\tilde \omega/\tilde\Delta$ and $u'$ is Re $u$.  The
self-consistency equation is then
\begin{eqnarray}
u &=& {\omega-\Sigma_0\over \Delta +\Sigma_1} =
{(\omega/\Delta)\sqrt{u^2-1} + i\,{\rm sgn}\,u'
\,\overline\Gamma
u\over \sqrt{u^2-1}+i\,{\rm sgn}\,u' \overline\Gamma},
\end{eqnarray}
with $\overline\Gamma=\Gamma/\Delta$, which has the solution
\begin{equation}
u = {\tilde \omega\over \tilde \Delta} = {\omega\over \Delta}.
\end{equation}
This means, as is well known, that the density of states
\begin{equation}\rho(\omega) = {\rm
Re}~ {\tilde \omega \, {\rm sgn}\,\omega\over \sqrt{\tilde\omega^2
-\Delta^2}} = {|\omega |\over \sqrt{\omega^2-\Delta^2}},
\end{equation}
is not changed by this type of impurity scattering and there is no
renormalization of the momentum-integrated thermodynamic
properties.  This is the essence of ``Anderson's theorem"
\cite{Andersontheorem}. On the other hand, one might expect that
the spectral weight $A(\k, \omega)$ for a quasiparticle of
momentum $\k$ should be broadened by disorder since the scattering
mixes different momentum states. This is the case in the normal
state where the spectral weight in the presence of impurity
scattering becomes a Lorentzian of width $2\Gamma$. To determine
what happens in the superconducting state we need the 11 component
of the Nambu Green's function
\begin{eqnarray} G_{11}(\k,\omega)&=&
\frac{\left(1+\frac{i\Gamma {\rm sgn}\,
\omega}{\sqrt{\omega^2-\Delta^2}}\right)
\omega+\epsilon_k}{\left(1+\frac{i\Gamma {\rm sgn}\,
\omega}{\sqrt{\omega^2-\Delta^2}}\right)^2
(\omega^2-\Delta^2)-\epsilon^2_k} \label{swaveG}
\end{eqnarray}

\begin{figure}
\leavevmode\includegraphics[width=3in]{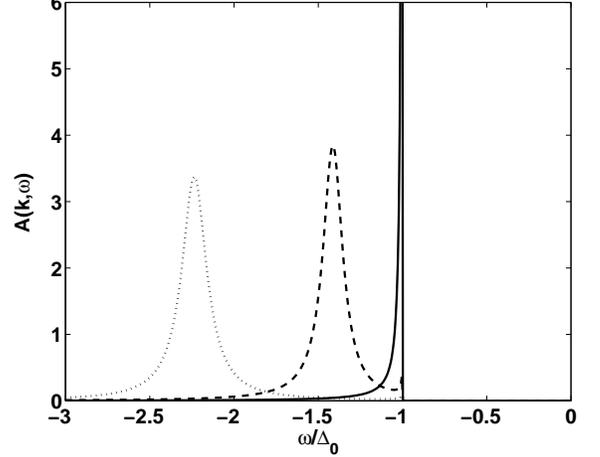}
 \caption{$A(\k,\omega)$ for an $s$-wave superconductor vs. $\omega/\Delta$ for
 $\Gamma/\Delta=0.1$ for 3 values (solid, dashed, dotted) of
 $\epsilon_\k/\Delta=0,-1,-2$.}
 \label{swaveAkw_vs_xi}
\end{figure}

\begin{figure}
\leavevmode
\includegraphics[width=\columnwidth]{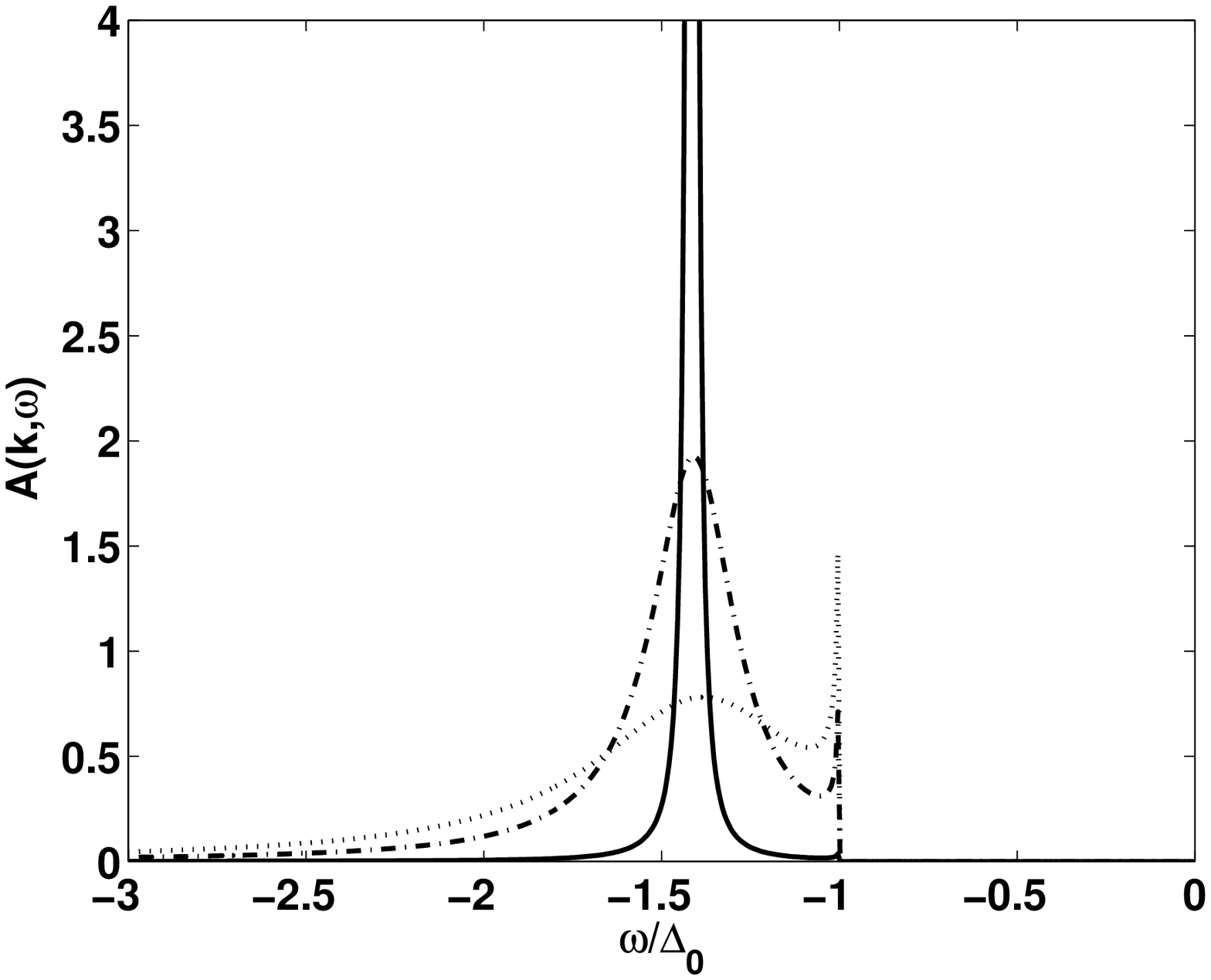}
 \label{swaveAkw_vs_gamma}
 \caption{$A(\k,\omega)$ for an $s$-wave superconductor vs.~$\omega/\Delta$ for
 $\epsilon_\k/\Delta=-1$ for 3 values (solid, dashed, dotted) of
 $\Gamma/\Delta=0.01$, 0.2, 0.5.}
\label{fig:swaveAkw_vs_gamma}
\end{figure}

In Figure \ref{swaveAkw_vs_xi}, we plot the spectral function
$A(\k, \omega)=-{\rm Im}\, G_{11}(\k, \omega)/\pi$ for $\omega <
0$ for various values if $k$ on and near the fermi surface.  Since
$i\Gamma/\sqrt{\omega^2-\Delta^2}$ is pure real for
$|\omega|/\Delta <1$, the spectral weight vanishes for energies
$|\omega|<\Delta$ the (unrenormalized) gap. However, as $\omega$
approaches $-\Delta$ from below, the spectral weight for $k=k_F$
diverges as
\begin{equation}
A(\k_F, \omega) \cong \frac{\Delta}{\pi\Gamma} \
\frac{1}{\sqrt{\omega^2-\Delta^2}} \label{aeight}
\end{equation}
As seen in Fig.~\ref{swaveAkw_vs_xi}, this square root singularity
gives way to a more symmetrically shaped dispersing quasiparticle
peak as $\k$ moves away from $\k_F$.  However, for $\k$ not too
far from the Fermi level, a residual  square root singularity at
$\omega = -\Delta$ remains.  In Figure \ref{fig:swaveAkw_vs_gamma}
which shows the dependence of the spectra on disorder, we see that
away from the Fermi level the strength of the structure at
$\omega=-\Delta$ increases with disorder.   In fact one can easily
show that when $|\epsilon_k|>\Gamma$, the spectral weight as
$\omega$ approaches $-\Delta$ varies as $A(\k, \omega) \simeq
\frac{\Gamma\Delta}{\pi\, \epsilon^2_k} \
\frac{1}{\sqrt{\omega^2-\Delta^2}}$. Thus in cleaner systems, the
anomalous peak at $\omega=-\Delta$ disappears and one has just the
expected quasiparticle peak at $\omega=-E_k$.

\begin{thebibliography}{99}
\bibitem{Wel95} B.~Wells, , Z. -X. Shen, A. Matsuura, D. M. King, M. A. Kastner,
M. Greven, and R. J. Birgeneau {\sl Phys.~Rev.~Lett.} {\bf 74},
964 (1995).

\bibitem{Ron98} F.~Ronning,  C. Kim, D. L. Feng, D. S. Marshall,
A. G. Loeser, L. L. Miller, J. N. Eckstein,
I. Bozovic, and Z.-X. Shen {\sl Science} {\bf 282}, 2067 (1998).

\bibitem{Val00} T.~Valla,  A. V. Fedorov, P. D. Johnson, Q. Li, G. D. Gu, and N. Koshizuka,
 {\sl Phys.~Rev.~Lett.} {\bf 85}, 828
(2000).

\bibitem{Kam00} A.~Kaminski,  J. Mesot, H. Fretwell, J. C. Campuzano, M. R. Norman,
M. Randeria, H. Ding, T. Sato, T. Takahashi, T. Mochiku, K.
Kadowaki, and H. Hoechst  {\sl Phys.~Rev.~Lett.} {\bf 84}, 1788
(2000).


\bibitem{DHS03} A.~Damascelli, Z.~Hussain, and Z.-X.~Shen,
{\sl Rev. Mod.~Phys.} {\bf 75}, 473 (2003).
\bibitem{AV00} E.~Abrahams and C.M.~Varma, {\sl Proc.~Nat'l Acad.~Sci.}
{\bf 97}, 5714 (2000).

\bibitem{Gia94} J. Giapintzakis, D. M. Ginsberg, M. A. Kirk and S. Ockers {\sl Phys.~Rev.~B} {\bf 50},
15967 (1994).

\bibitem{Tcsupress}  G.~Haran and A.D.S.~Nagi, {\sl Phys.~Rev.~B} 54, 15463-15467 (1996);
ibid, {\bf 58}, 12441 (1998);M. L. Kulic and O. V. Dolgov,  Phys.
Rev. B 60, 13062 (1999).


\bibitem{Kee01} H.-Y. Kee, {\sl Phys.~Rev.~B} {\bf 64}, 012506 (2001)


\bibitem{Fen01} D.L.~Feng,   N. P. Armitage, D. H. Lu, A. Damascelli,
J. P. Hu, P. Bogdanov, A. Lanzara, F. Ronning, K. M. Shen, H.
Eisaki, C. Kim, and Z.-X. Shen, J.-I. Shimoyama and K. Kishio ,
{\sl Phys.~Rev.~Lett.} {\bf 86}, 5550 (2001).

\bibitem{Chu01} Y.-D.~Chuang, A. D. Gromko, A. Fedorov, Y. Aiura, K. Oka,
Yoichi Ando, H. Eisaki, S. I. Uchida, and D. S. Dessau {\sl
Phys.~Rev.~Lett.} {\bf 87}, 117002 (2001).

\bibitem{Bog01} P.V.~Bogdanov,  A. Lanzara, X. J. Zhou, S. A. Kellar, D. L. Feng,
E. D. Lu, H. Eisaki, J.-I. Shimoyama, K. Kishio, Z. Hussain, and
Z. X. Shen {\sl Phys.~Rev.~B} {\bf 64}, 180505 (2001).

\bibitem{Kor02} A.A.~Kordyuk, S. V. Borisenko, T. K. Kim, K. A. Nenkov, M. Knupfer,
J. Fink, M. S. Golden, H. Berger, and R. Follath {\sl
Phys.~Rev.~Lett.} {\bf 89}, 77003 (2002).

\bibitem{Kam04} A.~Kaminski {\it et.~al}; cond-mat/0404385.

\bibitem{Eisaki}H. Eisaki, N. Kaneko, D. L. Feng, A. Damascelli, P. K. Mang, K. M. Shen, Z.-X. Shen, and M. Greven
Phys. Rev. B 69, 064512 (2004).








\bibitem{ZAH}   Lingyin Zhu, W.A. Atkinson, and P. J.
Hirschfeld,   Phys. Rev. B 69, 060503 (2004).

\bibitem{McE03} J. E. Hoffman {\em et al.}, Science {\bf 295}, 466
(2002); K.~McElroy {\it et.~al}, {\sl Nature} {\bf 422}, 592
(2003).
 \bibitem{Mar03}R.S.~Markiewicz;
cond-mat/0309254.
















\bibitem{DHS01} D.Duffy, D.J. Scalapino, and P.J. Hirschfeld, Phys. Rev. B 64,
224522 (2001).



\bibitem{Andersontheorem} P. W. Anderson, {Phys. Rev. Lett.} \textbf{\ 3},
328 (1959).



\bibitem{AG} Abrikosov-Gorkov, Zh. Eksperim. i Teor. Fiz. 39, 1781 (1960) [English translation
Sov. Phys. -- JETP 12, 1243 (1961).




\bibitem{AZH03} W.A.
Atkinson, P.J. Hirschfeld and L. Zhu, Phys. Rev. B 68, 054501
(2003).



\bibitem{Quinlan94}S.~M.~Quinlan, D.~J.~Scalapino, and N.~Bulut,
{\sl Phys.~Rev.~B}, {\bf 49}, 1470 (1994).



\bibitem{Yashenkin}M. L. Titov, A.G. Yashenkin, and D.N. Aristov,
Phys. Rev. B52, 10626 (1995).

\bibitem{QHS96}S.M. Quinlan,
P.J. Hirschfeld, and D.J. Scalapino, Phys. Rev. B. 53, 8575
(1996).



%
%









\end{thebibliography}
\end{document}